\documentclass[a4paper]{jpconf}
\usepackage{graphicx}

\begin{document}
\title{Prospects for observing extreme-mass-ratio inspirals with LISA}

\author{Jonathan R Gair$^{1,\dag}$, Stanislav Babak$^2$, Alberto Sesana$^3$, Pau Amaro-Seoane$^{4,5,6,7}$, Enrico Barausse$^{8,9}$, Christopher P L Berry$^3$, Emanuele Berti$^{10,11}$ and Carlos Sopuerta$^4$}

\address{$^1$School of Mathematics, University of Edinburgh, The King's Buildings,
Peter Guthrie Tait Road, Edinburgh, EH9 3FD, UK}
\address{$^2$Max Planck Institute for Gravitational Physics, Albert Einstein Institute, Am M\"{u}hlenberg 1, 14476 Golm, Germany}
\address{$^3$School of Physics and Astronomy, University of Birmingham, Edgbaston, Birmingham B15 2TT, UK}
\address{$^4$Institut de Ci\`{e}ncies de l'Espai (CSIC-IEEC), Campus UAB, Carrer de Can Magrans s/n, 08193 Cerdanyola del Vall\`{e}s, Spain}
\address{$^5$Kavli Institute for Astronomy and Astrophysics, Beijing 100871, China}
\address{$^6$Institute of Applied Mathematics, Academy of Mathematics and Systems Science, CAS, Beijing 100190, China}
\address{$^7$Zentrum f{\"u}r Astronomie und Astrophysik, TU Berlin, Hardenbergstra{\ss}e 36, 10623 Berlin, Germany}
\address{$^8$Sorbonne Universit\'{e}s, UPMC Univesit\'{e} Paris 6, UMR 7095, Institut d'Astrophysique de Paris, 98 bis Bd Arago, 75014 Paris, France}
\address{$^9$CNRS, UMR 7095, Institut d'Astrophysique de Paris, 98 bis Bd Arago, 75014 Paris, France}
\address{$^{10}$Department of Physics and Astronomy, The University of Mississippi, University, MS 38677, USA}
\address{$^{11}$CENTRA, Departamento de F\'isica, Instituto SuperiorT\'ecnico, Universidade de Lisboa, Avenida Rovisco Pais 1, 1049 Lisboa, Portugal}

\ead{$^\dag$j.gair@ed.ac.uk}

\begin{abstract}
One of the key astrophysical sources for the Laser Interferometer Space Antenna (LISA) are the inspirals of stellar-origin compact objects into massive black holes in the centres of galaxies. These extreme-mass-ratio inspirals (EMRIs) have great potential for astrophysics, cosmology and fundamental physics. In this paper we describe the likely numbers and properties of EMRI events that LISA will observe. We present the first results computed for the $2.5~\mathrm{Gm}$ interferometer that was the new baseline mission submitted in January 2017 in response to the ESA L3 mission call. In addition, we attempt to quantify the astrophysical uncertainties in EMRI event rate estimates by considering a range of different models for the astrophysical population. We present both likely event rates and estimates for the precision with which the parameters of the observed sources could be measured. We finish by discussing the implications of these results for science using EMRIs.
\end{abstract}

\section{Introduction}
The first gravitational wave (GW) detections by LIGO~\cite{GW150914,GW151226} in 2015 began the new field of GW astronomy. These events indicated that the ground-based interferometers will see many more systems over the coming decades~\cite{O1Summary}. However, seismic noise limits the sensitivity of ground-based detectors to frequencies above $\sim1~\mathrm{Hz}$ and hence to systems with total mass no more than a few hundred solar masses. Observational evidence indicates that the centres of most galaxies contain massive black holes (MBHs), with masses between a few tens of thousands and a few billion solar masses~\cite{croton}. Mergers involving MBHs are powerful sources of GWs, but these GWs can only be observed from space. ESA selected GW detection from space as the theme to be addressed by the L3 mission in the Cosmic Vision programme, scheduled for launch in 2034~\cite{tgu}. The technology for a space-based interferometer was successfully demonstrated in 2016 by the LISA Pathfinder satellite~\cite{LPFResults}, which paved the way for an ESA call for L3 mission proposals that closed in early 2017. Given the success of LISA Pathfinder, it is highly likely that the successful proposal will be the Laser Interferometer Space Antenna (LISA)~\cite{L3missprop}.

LISA will have sensitivity to GWs in the millihertz band, which are generated by merging systems having total mass in the range $\sim10^4$--$10^7M_\odot$. LISA is expected to observe a variety of sources, including hour-period compact binaries (primarily white dwarf--white dwarf binaries) in the Milky Way, the early inspiral phase of some of the heavier stellar-origin black hole binaries that LIGO has shown to exist~\cite{LIGOLISABinaries}, and possibly cosmological sources such as GWs from cosmic string cusps or a stochastic background generated by phase transitions in the early Universe. However, the primary source for LISA are the MBHs in the centres of galaxies. These MBHs will generate GWs either when they merge with other MBHs, which is expected to occur following mergers between their host galaxies, or when they merge with much smaller compact objects formed from stars. These latter systems are called extreme-mass-ratio inspirals (EMRIs), because of the large difference in mass between the two objects involved. Galactic MBHs are typically surrounded by a stellar cluster, and EMRIs occur when compact objects formed as the end point of the evolution of stars in that cluster are captured and gradually fall into the central MBH. EMRIs occur over long timescales, meaning that $\sim 10^4$--$10^5$ cycles of gravitational radiation are generated while the compact object is in the strong gravitational field region close to the central MBH. In addition, the process by which EMRIs begin, through a scattering capture of the compact object by the MBH, tends to lead to high initial eccentricities, with the result that EMRI orbits are expected to retain significant eccentricity and non-equatorial inclinations when they are radiating in the LISA band. These properties lead to a great richness in EMRI gravitational waveforms, created by a superposition of the three fundamental frequencies --- the orbital frequency, the perihelion precession frequency and the frequency of precession of the orbital plane. This complexity contains a great deal of information that can be used for science. EMRI GWs can be used both to measure the parameters of the system to high precision and to detect small deviations in the waveforms from the predictions of general relativity that are indicative of a breakdown of the theory. EMRIs thus have tremendous potential for astrophysics, cosmology and fundamental physics; this will be discussed again at the end of this article, but we refer the reader to~\cite{EMRIReview,TestGRLivRev} for a more comprehensive discussion.

In this article we will compute the likely numbers of EMRI events and the precision with which LISA will be able to measure their parameters. There have been previous studies computing expected numbers of EMRI events~\cite{JG09}. However, the astrophysical model employed in those calculations was a combination of simple power laws and no attempt was made to quantify the uncertainties in that model. EMRI parameter-estimation studies have also been carried out~\cite{AK,EliuNK}, but only for a small sample of representative cases and not for a full astrophysical population. We address both of these shortcomings in the current article. We compute event rates for several different astrophysical models that cover the range that is plausible given current observational constraints and compute estimates of the parameter-estimation accuracies for all the events in each population. In addition, we compute these results for the first time for a $2.5~\mathrm{Gm}$ LISA detector with six laser links, which was proposed as the new mission baseline in the response to the ESA call in January 2017~\cite{L3missprop}.

This article is a summary of a paper that has recently been submitted~\cite{PRDPaper} and we refer the interested reader to that paper for more details of the calculations and final results. This paper is organised as follows. In Section~\ref{sec:model}, we describe the models we employ for the detector sensitivity, the EMRI waveforms and the astrophysical population of EMRI sources. In Section~\ref{sec:results} we present a comparison of the likely numbers of EMRI events that will be observed under each model and the parameter-estimation precision. In Section~\ref{sec:science} we describe the implications of these results for science with EMRIs, before finishing in Section~\ref{sec:conc} with a short summary of conclusions.

\section{Model}
\label{sec:model}
To estimate the detectability of EMRIs by LISA, we must model three different things --- the intrinsic astrophysical population of EMRI events; the GWs generated by EMRI events; and the sensitivity of our detector to those GWs. We describe the models we use for each of these in this section. 

\subsection{Intrinsic astrophysical population}
Our astrophysical population model includes models for the MBH mass and spin distribution, the intrinsic rate of EMRIs occurring in systems containing black holes with particular properties and the properties of the compact object in the EMRI.

{\bf MBH mass distribution} Previous estimates~\cite{JG09} of EMRI rates have used a simple power law MBH mass function of the form
\begin{equation}
\frac{{\rm d}n}{{\rm d}\log M} = A \left(\frac{M}{3\times 10^6 M_\odot} \right)^\alpha \mathrm{Mpc}^{-3}
\end{equation}
which is known to be a good fit to the observed MBH population in the relevant range~\cite{GTV}. Values of $A=0.002$ and $-0.3 \leq \alpha \leq 0.3$ are usually assumed. A more sophisticated model that self-consistently evolves the MBH population is now available~\cite{Barausse12}, and has been used to predict the expected number of MBH binary mergers that will be observed by LISA~\cite{GOATMBH}. We will denote this self-consistent MBH model as ``B12''. This model generates mass functions that are roughly consistent with the above power-law form, with $A=0.005$ and $\alpha=-0.3$. To obtain a lower bound on the EMRI rate we will also consider a strict power-law MBH mass function with $A=0.002$ and $\alpha=0.3$. We will denote this model as ``G10''.

During a major merger, we expect the stellar cusp around the MBH to be disrupted. The cusp will regrow, but this takes a quarter of the relaxation time for a Milky Way-like galaxy~\cite{PretoAS10}, during which the standard EMRI picture does not apply. We include this effect in our model by assuming an MBH is unavailable as a host for EMRIs for a time
\begin{equation}
t_{\rm cusp}= T_0 (M/10^6M_\odot)^{1.19} q^{0.35}\,\mathrm{Gyr}
\label{eq:cuspregrow}
\end{equation}
following a merger with total binary mass $M$ and mass ratio $q$. The pre-factor $T_0$ depends on the assumed $M$--$\sigma$ relation, which is discussed below. We cannot apply this correction for the ``G10'' MBH population, as we do not track mergers in that case.

{\bf MBH spin distribution} The ``B12'' model also tracks MBH spins. These are driven partially by mergers but primarily by accretion. We denote the MBH spin distribution computed in ``B12'' by ``a98'' as it predicts~\cite{SBRD14} that most MBHs will be spinning close to the maximal imposed limit, i.e., with $a\approx 0.98$. To understand the significance of this assumption we also consider a ``flat'' model with spins uniformly distributed in the range $a \in [0,0.98]$, and a conservative ``a0'' model in which all spins are set to $a=0$.

{\bf EMRI rate per black hole} We base our EMRI rate per MBH on~\cite{ASPreto11}
\begin{equation}
R_0 = 300 \left( \frac{M}{10^6M_\odot} \right)^{-0.19} \mathrm{Gyr}^{-1}.
\label{eq:rateref}
\end{equation}
If this rate was maintained over the age of the Universe, it would overgrow lower mass black holes, particularly when we account for the fact that for every EMRI there will be $N_\mathrm{p} \sim 1$--$100$ ``failed EMRIs'' that directly plunge~\cite{Merritt15}. Therefore, we add two corrections. First, we compute $\Gamma = \min\{t_\mathrm{d}/t_\mathrm{relax},1\}$, where $t_\mathrm{relax}$ is the relaxation time in the MBH stellar cusp and $t_\mathrm{d} = 0.06M/[m R_0 (1+N_\mathrm{p})]$ is the time it would take to deplete the stellar cusp of compact objects through EMRIs and direct plunges. Second, we compute $\kappa = \min\{M/({\rm e}\, \Gamma \Delta M),1\}$, where $\Delta M$ is the total mass that would be accreted by the MBH over its evolution history according to~(\ref{eq:rateref}). The final EMRI rate is taken to be $\kappa \Gamma R_0$. The $\Gamma$ factor corrects for the time taken for the stellar loss cone to be repopulated by diffusion, while the $\kappa$ factor imposes a limit (of one e-fold) on the total mass that an MBH can gain by compact object accretion.

{\bf Black hole $M$--$\sigma$ relation} The factor $T_0$ in~(\ref{eq:cuspregrow}) and the stellar relaxation time depend on the velocity dispersion $\sigma$ in the stellar cluster. This can be determined from the black hole mass via the $M$--$\sigma$ relation. We use the model of \cite{Gult09} as our reference model ``Gultekin09'', which predicts $T_0\sim 6$, plus an optimistic case ``GrahamScott13''~\cite{GrahamScott13}, which predicts $T_0 \sim 2$, and a pessimistic case ``KormendyHo13''~\cite{KormendyHo13}, which predicts $T_0 \sim 10$.

{\bf Compact object properties} EMRI rates are expected to be dominated by inspirals of black holes, as these are visible to much greater distances and are intrinsically enhanced by mass segregation~\cite{EMRIReview,JG09}. Typically the mass, $m$, of the inspiralling object has been taken to be fixed at $m=10M_\odot$. Given recent LIGO results~\cite{GW150914,O1Summary} we will consider both this standard case and one in which all EMRIs are assumed to have $m=30M_\odot$. We take the eccentricity of the EMRI orbit at plunge to be uniformly distributed in the range $[0,0.2]$. EMRIs on prograde orbits into more rapidly spinning MBHs can be more easily captured as inspirals rather than direct plunges. We account for this by enhancing the EMRI rate by a factor $W(a)$ that is a function of the MBH spin $a$, and by using the inclination distribution $p(\iota)\propto \sin(\iota) [W(a,\iota)]^{-0.83}$~\cite{ASSF13}.

\begin{table}
\begin{center}
\begin{tabular}{cccccccc}%\hline
ID&MF&MBH spin&Erosion&$M$--$\sigma$ relation&$N_\mathrm{p}$&CO mass $[M_\odot]$&EMRI rate [$\mathrm{yr}^{-1}$]\\\hline
M1&B12&a98&yes&Gultekin09&10&10&1600\\
M2&B12&a98&yes&KormendyHo13&10&10&1400\\
M3&B12&a98&yes&GrahamScott13&10&10&2770\\
M4&B12&a98&yes&Gultekin09&10&30&520\\
M5&G10&a98&no&Gultekin09&10&10&140\\
M6&B12&a98&no&Gultekin09&10&10&2080\\
M7&B12&a98&yes&Gultekin09&10&10&15800\\
M8&B12&a98&yes&Gultekin09&100&10&180\\
M9&B12&flat&yes&Gultekin09&10&10&1530\\
M10&B12&a0&yes&Gultekin09&10&10&1520\\
M11&G10&a0&no&Gultekin09&100&10&13\\
M12&B12&a98&no&Gultekin09&10&10&20000\\
\end{tabular}
\end{center}
\caption{\label{tab:modelsumm}Summary of the twelve models used in this study. The columns indicate, respectively, the model identifier, the mass function prescription, the model for MBH spins, whether cusp erosion following mergers is included or not, the model used for the $M$-$\sigma$ relation, the number of direct plunges per EMRI, the assumed mass of the compact object in all EMRIs, and the resulting astrophysical rate of EMRI events in the Universe at redshift $z < 4.5$.}
\end{table}

Overall, we use $12$ different models that are constructed using different combinations of the above ingredients. The models are summarised in Table~\ref{tab:modelsumm}. M1 is our reference model. M10 and M11 will be used only to estimate event rates, as degeneracies present for non-spinning EMRIs make parameter-estimation calculations based on the Fisher matrix more challenging. Full parameter-estimation results will appear in the longer journal article based on this work.

\subsection{EMRI gravitational waveforms}
The extreme-mass-ratio in an EMRI system means that the emitted GWs can be computed accurately using black hole perturbation theory. This involves calculating the self-force acting on the inspiralling object. We refer the reader to~\cite{poissonLRR} for details on this approach. Self-force calculations have not yet been completed at the necessary order for objects on arbitrary orbits about spinning black holes, and those calculations that do exist for simpler cases are computationally intensive and therefore not well suited to studies of populations such as this one. As an alternative, there are two different families of kludge waveforms available. These are approximate models that have been constructed to include the most important qualitative features of EMRI waveforms, while not necessarily being quantitatively accurate. The analytic kludge (AK) model~\cite{AK} constructs an EMRI waveform starting from the simple GW emission from a Keplerian orbit, and then imposing precession and evolution of the orbit onto the system. The precessional and evolution effects are governed by low-order post-Newtonian equations, which are accurate in the weak-field, low-velocity regime. The AK model is cheap to generate, but it rapidly goes out of phase with more accurate waveforms. The numerical kludge (NK) model~\cite{GG06,NK07} uses Kerr geodesics as the basis for constructing EMRI orbits, adding inspiral determined by evolution equations that are based on post-Newtonian expansions, but augmented by fits to perturbation theory results. The NK waveform is generated from the trajectory using a flat-space emission formula~\cite{NK07}. NK waveforms are much more faithful to accurate inspirals than AK waveforms, staying in phase down to the last stages of inspiral~\cite{NK07}; however, they are more computationally expensive.

For the purposes of this study we need to be able to generate large numbers of gravitational waveforms to determine the detectability and parameter-estimation precision for astrophysical populations of EMRI events. We therefore use the computationally cheapest of the three models, the AK. While these waveforms are not faithful to the true EMRI signals produced in nature, it is believed that they capture the richness of real EMRI gravitational waveforms and therefore will provide an accurate guide to signal to noise ratios and parameter-estimation precisions. To attempt to characterise the uncertainty from the waveform choice, we consider two different versions of the AK. In the original AK model~\cite{AK}, the waveforms are terminated when the object reaches the Schwazrschild innermost stable orbit (ISO). This is inaccurate for prograde inspirals into spinning black holes, which can get much closer to the central black hole. As the spins of the black holes in our population tend to be quite high, this is likely to significantly underestimate the detectability of many events. A simple fix is to continue the inspiral until the EMRI reaches the Kerr ISO. As the post-Newtonian expressions on which the AK model is built break down in the strong-field regime at the end of the inspiral, this continuation is not going to be accurate and will most likely lead to an overestimate of detectability. Therefore, we expect the two sets of results to bracket the true answer.

\subsection{LISA sensitivity to EMRIs}
An EMRI will be detectable by LISA if it is sufficiently loud. This can be characterised by a requirement on the matched filtering signal-to-noise ratio (SNR) $\rho$, which the source has in the detector data stream
\begin{equation}
\rho^2 = \langle h | h \rangle, \quad \mbox{where } \langle a|b \rangle = 2 \int_0^\infty \frac{\tilde{a}(f) \tilde{b}^*(f)+\tilde{a}^*(f) \tilde{b}(f)}{S_n(f)} \mathrm{d}f.
\end{equation}
This expression depends on the final detector configuration through $S_n(f)$, the power spectral density of noise in the detector. We use the LISA noise model described in~\cite{L3missprop}
\begin{equation}
S_n(f) = \frac{20}{3} \left[ L^2 + \left(\frac{2f}{0.41c} \right)^2\right] \left[ 4 S_n^\mathrm{acc}(f) + 2 S_n^\mathrm{loc}(f) +S_n^\mathrm{sn}(f) +S_n^\mathrm{omn}(f)\right],
\end{equation}
where $S_n^\mathrm{acc}(f)=\left\{9\times10^{-30}+3.24 \times 10^{-28}\left[ (3\times10^{-5} \mathrm{Hz}/f)^{10} + (10^{-4} \mathrm{Hz}/f)^2\right]\right\}/(2\pi f)^4~\mathrm{m^2\,Hz^{-1}}$, $S_n^\mathrm{loc}(f)=2.89\times10^{-24}~\mathrm{m^2\,Hz^{-1}}$, $S_n^\mathrm{sn}(f)=7.92\times10^{-23}~\mathrm{m^2\,Hz^{-1}}$ and $S_n^\mathrm{omn}(f)=4\times10^{-24}~\mathrm{m^2\,Hz^{-1}}$. We set the detector arm length to $L=2.5$Gm. The complexities of EMRI data analysis mean that a relatively high SNR will be required for confident detection. The value $\rho_\mathrm{ thresh} = 30$ was historically assumed~\cite{EMRIrate04}, but results from the Mock LISA Data Challenge suggest that EMRIs with SNR as low as $\rho_\mathrm{thresh} = 20$ could be identified~\cite{MLDC3}, albeit under somewhat idealised conditions. We will present results for both thresholds.

We assume conservatively that the duration of the LISA mission is two years and that any EMRIs that are detected must plunge during the mission lifetime. EMRIs accumulate SNR gradually over the whole inspiral, so LISA will also observe systems that are in the final stages of inspiral but do not reach plunge over the mission duration. The event rates reported here are therefore conservative in this regard. We use two years as this was the default used in the suite of recent studies connected to possible LISA configurations, such as~\cite{GOATMBH}. The recent mission proposal~\cite{L3missprop} uses a longer nominal mission duration of four years. To extrapolate to longer mission durations we can multiply by the ratio of the mission duration to two years, so a four year mission will have approximately twice as many EMRI events as reported here. Again, this will be conservative, as some EMRI signals that do not accumulate enough SNR over two years will pass the SNR threshold when integrated over a longer observation.

To evaluate parameter-estimation precision we use the Fisher matrix, defined by
\begin{equation}
\Gamma_{ij} = \left\langle \frac{\partial h}{\partial \lambda_i} \bigg{|} \frac{\partial h}{\partial \lambda_j} \right \rangle\,,
\end{equation}
where $\lambda^i$ denotes the waveform model parameters. An estimate of the precision of measurement is given by $\Delta \lambda_i ^2 = (\Gamma^{-1})_{ii}$ (no sum over $i$).

\section{Results}
\label{sec:results}
\subsection{Event rates}
Table~\ref{tab:AKrates} indicates the number of EMRI events that would be detected over the two year mission lifetime for each model, using SNRs computed for AK waveforms with the standard Schwarzschild plunge condition. For the reference model we would expect to observe several hundred events with SNR above $20$, of which approximately one third would have SNR above $30$. The number of observed EMRIs can be no more than a factor of $10$ larger (M12), if we assume that MBHs are always available as EMRI hosts and MBHs gain no mass from direct plunges of compact objects. The EMRI rate could be as low as $0$, if we assume a mass function that falls steeply toward lower MBH masses and that MBHs have small spins (M11). That model is overly pessimistic, and M5 and M8 probably give more realistic lower bounds, which are a factor of $10$ smaller. The majority of the models predict several hundred EMRI events, so this figure appears to be fairly robust to astrophysical uncertainties. 

If the inspiralling compact objects in EMRIs tend to be more massive (M4), we see a comparable number of events. While such EMRIs can be seen to larger redshift, this is compensated by a decrease in the intrinsic rate to prevent MBH overgrowth. Table~\ref{tab:NKrates} shows the corresponding results computed using the Kerr plunge criterion to terminate the EMRI. These rates are somewhat higher, as expected, but only by a factor $\sim 1$--$2$.

In our model we have prevented MBH overgrowth through compact object accretion by imposing a constraint on the EMRI rate. An alternative solution is that the number of light MBHs is significantly reduced, because these MBHs grow rapidly from EMRI consumption. Results in the table are divided into mass bins and we see that between $20\%$ and $60\%$ of the events have MBH mass greater than $10^{5.5}M_\odot$, so even if no lighter MBHs exist we should see approximately one hundred events in the reference model. The rate drops significantly, by a factor of $40$ or more, if we impose a cut off at $10^6M_\odot$, when using the Schwarzschild plunge condition. However, a higher proportion of the events are heavier mass when we use the Kerr plunge condition. This is because in that model the EMRI can get closer to the MBH, shifting the GW emission to higher frequencies and hence providing sensitivity to heavier MBHs. The prospects for significant numbers of EMRI detections are therefore good even if the number of lower mass MBHs is significantly depleted.

%\begin{table}
%\begin{center}
%\begin{tabular}{c|c|c|c|c|c|c|c|c}
%&\multicolumn{4}{|c|}{Number of events with SNR $>20$ and}&\multicolumn{4}{|c|}{Number of events with SNR $>30$ and}\\
%Model&$M_{10}< 5$&$5 < M_{10} < 5.5$&$5.5 < M_{10} < 6$&$6 < M_{10}$&$M_{10} < 5$&$5 < M_{10} < 5.5$&$5.5 < M_{10} < 6$&$6 < M_{10}$\\\hline
%\end{tabular}
%\end{center}
%\caption{\label{tab:AKrates}$M_{10}=\log_{10}(M/M_\odot)$}
%\end{table}

\begin{table}
\begin{center}
\begin{tabular}{c|c|c|c|c|c|}
&\multicolumn{4}{|c|}{Number of events in mass range}&\\
Model&$M_{10}< 5$&$5 < M_{10} < 5.5$&$5.5 < M_{10} < 6$&$6 < M_{10}$&Total\\\hline
M1&20 (10)&240 (60)&110 (50)&10 (0)&380 (130)\\
M2&30 (10)&190 (50)&70 (30)&0 (0)&290 (90)\\
M3&20 (0)&310 (90)&510 (220)&40 (20)&880 (340)\\
M4&70 (20)&280 (130)&80 (50)&0 (0)&440 (200)\\
M5&0 (0)&10 (0)&20 (10)&0 (0)&30 (10)\\
M6&20 (0)&270 (70)&210 (90)&20 (10)&520 (180)\\
M7&230 (50)&2190 (600)&1040 (480)&60 (40)&3530 (1170)\\
M8&0 (0)&30 (10)&10 (10)&0 (0)&50 (10)\\
M9&20 (10)&210 (60)&110 (50)&10 (10)&350 (130)\\
M10&30 (10)&240 (70)&100 (40)&10 (10)&370 (130)\\
M11&0 (0)&0 (0)&1 (0)&0 (0)&1 (0)\\
M12&230 (50)&2420 (670)&1730 (730)&180 (110)&4560 (1560)\\
\end{tabular}
\end{center}
\caption{\label{tab:AKrates}Number of events detected in each mass range, and total number of events, for each model. Mass ranges are indicated in terms of $M_{10}=\log_{10}(M/M_\odot)$. The primary values in each cell assume an SNR threshold of $20$ is required for detection, while the bracketed numbers given the corresponding results for an SNR threshold of $30$. SNRs are computed using the AK model with the Schwarzschild plunge condition. All numbers are rounded to the nearest 10 apart from the M11 results which are rounded to the nearest $1$.}
\end{table}

\begin{table}
\begin{center}
\begin{tabular}{c|c|c|c|c|c|}
&\multicolumn{4}{|c|}{Number of events in mass range}&\\
Model&$M_{10}< 5$&$5 < M_{10} < 5.5$&$5.5 < M_{10} < 6$&$6 < M_{10}$&Total\\\hline
M1&20 (0)&260 (60)&230 (100)&80 (60)&590 (230)\\
M2&20 (0)&210 (50)&160 (70)&50 (40)&440 (160)\\
M3&10 (0)&360 (90)&1000 (470)&240 (180)&1620 (750)\\
M4&50 (10)&300 (150)&140 (100)&30 (30)&520 (280)\\
M5&0 (0)&10 (0)&40 (20)&40 (30)&90 (50)\\
M6&20 (0)&300 (80)&430 (200)&200 (150)&960 (440)\\
M7&190 (40)&2390 (600)&2110 (930)&730 (510)&5420 (2090)\\
M8&0 (0)&30 (10)&30 (10)&10 (10)&70 (30)\\
M9&20 (0)&230 (60)&160 (70)&30 (20)&430 (160)\\
M10&30 (10)&240 (70)&100 (40)&10 (10)&370 (130)\\
M11&0 (0)&0 (0)&1 (0)&0 (0)&1 (0)\\
M12&190 (40)&2700 (680)&3710 (1690)&1830 (1380)&8440 (3790)\\
\end{tabular}
\end{center}
\caption{\label{tab:NKrates}As Table~\ref{tab:AKrates}, but now with SNRs computed using the AK model with the Kerr plunge condition.}
\end{table}

Table~\ref{tab:config} shows how the number of events detected depends on the configuration of the detector for the reference model M1. If the final LISA configuration is similar to the $1~\mathrm{Gm}$, four-link NGO model used in~\cite{tgu}, the number of events would be about a factor of $10$ smaller. If a more sensitive $5~\mathrm{Gm}$ configuration was launched, then the event rate could be increased by a factor of about three. The change in the number of detected events is similar for the other astrophysical models, except for the high mass compact object model (M4), for which the decrease in number of events going to the $1~\mathrm{Gm}$ configuration is only a factor of $\sim5$, and the increase going to the $5~\mathrm{Gm}$ configuration is only a factor of $\sim1.5$. This difference arises because the EMRIs in M4 are intrinsically louder and so all detector configurations can detect most of the EMRI events out to moderate redshift.

\begin{table}
\begin{center}
\begin{tabular}{cc|cc|cc}
\multicolumn{6}{c}{Armlength}\\
\multicolumn{2}{c|}{$1~\mathrm{Gm}$}&\multicolumn{2}{|c|}{$2.5~\mathrm{Gm}$}&\multicolumn{2}{|c}{$5~\mathrm{Gm}$}\\
4 links&6 links&4 links&6 links&4 links&6 links\\\hline
0.1&0.2&0.4&1&2&3
\end{tabular}
\end{center}
\caption{\label{tab:config}Variation in the number of observed EMRI events with detector configuration. Numbers are computed for the reference model, M1, and are expressed relative to the number detected for the new baseline $2.5~\mathrm{Gm}$, six-link configuration.}
\end{table}

\subsection{Parameter-estimation accuracy}
\begin{figure}
\begin{tabular}{cc}
\includegraphics[width=0.45\textwidth]{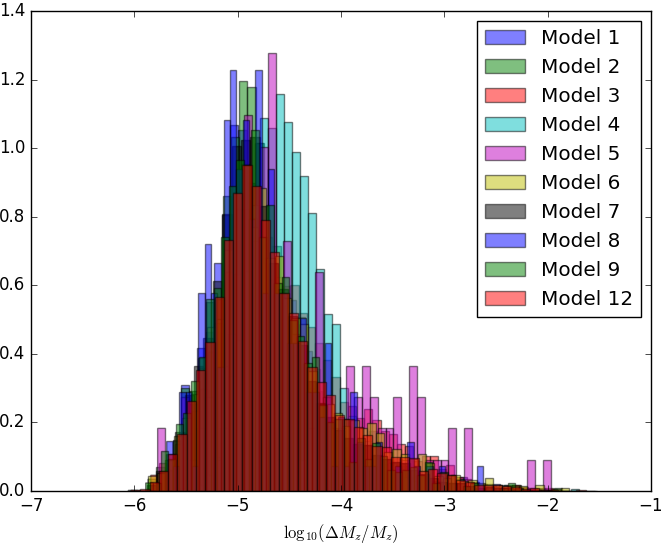}&
\includegraphics[width=0.45\textwidth]{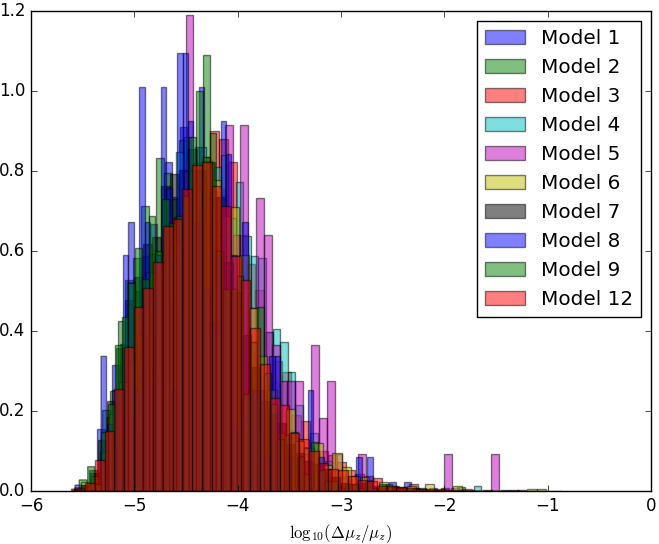}\\
\includegraphics[width=0.45\textwidth]{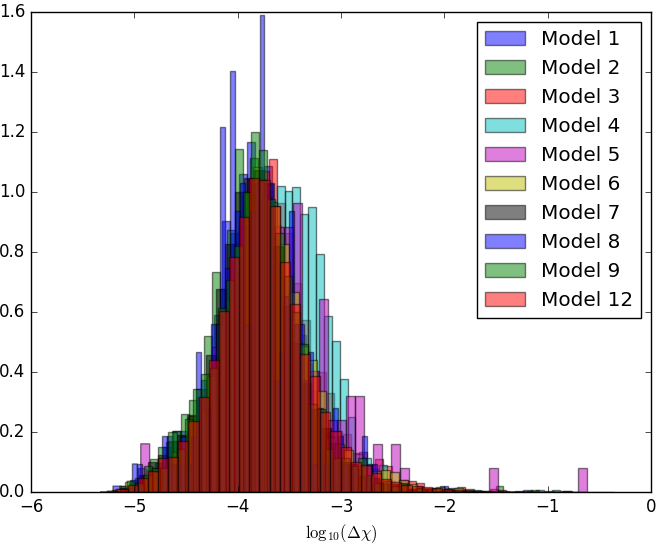}&
\includegraphics[width=0.45\textwidth]{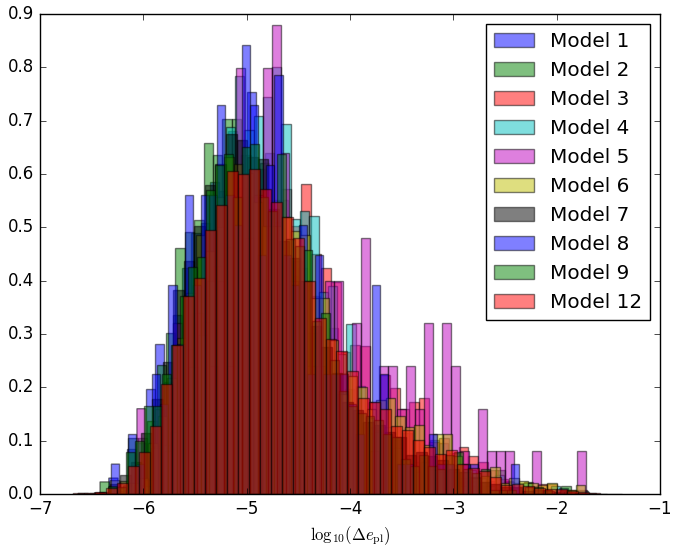}
\end{tabular}
\caption{\label{fig:intrinsPE}Expected parameter-estimation accuracies for redshifted MBH mass (top left), redshifted compact object mass (top right), MBH spin (bottom left) and eccentricity at plunge (bottom right).}
\end{figure}

\begin{figure}
\begin{tabular}{cc}
\includegraphics[width=0.45\textwidth]{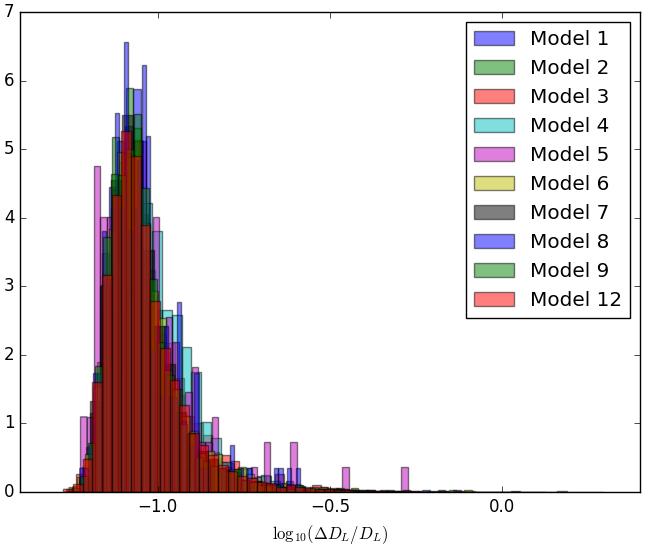}&
\includegraphics[width=0.45\textwidth]{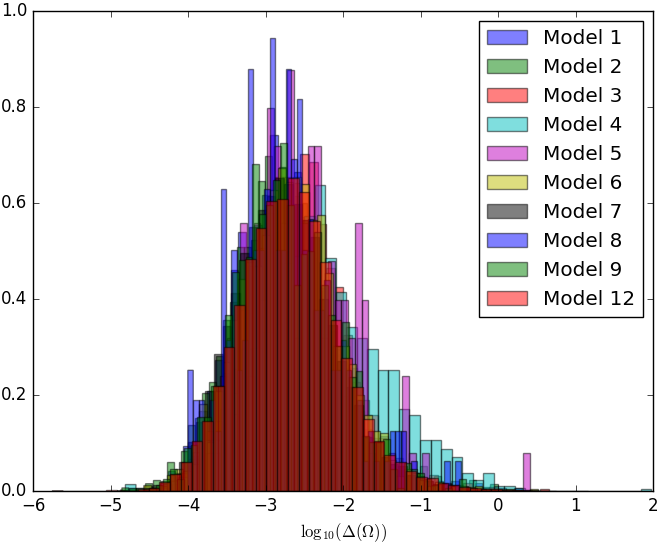}
\end{tabular}
\caption{\label{fig:extrinsPE}Expected parameter-estimation accuracies for luminosity distance (left) and sky position (right).}
\end{figure}

Figure~\ref{fig:intrinsPE} shows the distribution of parameter-estimation precisions that we would expect for the set of EMRIs observed in each population model. We present results for the redshifted masses of the MBH and compact object, for the spin of the MBH and for the eccentricity at plunge. We have normalised results to a fixed SNR of $20$ (accuracies scale as $\Delta X \propto 1/\rho$) and we see that after doing that the precision is essentially independent of the population model. The distribution of SNRs is determined by the spatial distribution of sources, which is uniform in comoving volume and hence also similar in each model. However, the models that predict larger numbers of events will have some sources at high SNRs which will have correspondingly more precise parameter determinations. We see that we would expect to measure the redshifted masses of the components and the orbital characteristics such as eccentricity to precisions of $\sim 10^{-6}$--$10^{-3}$, while also measuring MBH spins to $\sim10^{-5}$--$10^{-3}$. Figure~\ref{fig:extrinsPE} shows corresponding results for EMRI measurements of source distances and sky positions. Luminosity distance should typically be determined to a few to a few tens of percent, and sky location should be measured to about a thousandth of a steradian, which is a few square degrees. Sky position will be slightly less well measured for EMRIs with larger mass inspiraling objects, since such sources are observable for a smaller fraction of a LISA orbit. The luminosity distance accuracy is also approximately the accuracy with which the intrinsic masses of the sources will be determined, since the redshift will be inferred from the luminosity distance. We would therefore expect to determine the intrinsic MBH mass to better than $10\%$ in the majority of cases.

These results were computed using the Schwarzschild plunge condition. Results computed with the Kerr plunge condition are typically a factor of $\sim10$ better for the intrinsic parameters (and comparable for the extrinsic parameters), so these results are most likely conservative.

\section{Implications for EMRI science}
\label{sec:science}

\subsection{Astrophysics}
In~\cite{GTV} it was shown that the classic $5~\mathrm{Gm}$ LISA configuration would be able to measure the slope of the mass function of MBHs in the LISA range to a precision of $\sim \pm0.3$ if $10$ EMRI events were observed. This is the precision with which the mass function is currently known. This conclusion is robust to assumptions about the configuration of the detector. In the astrophysical models presented here, we expect more than $10$ EMRI events in two years for all but the most pessimistic model. In most cases, the number of EMRIs expected is several hundreds. The  precision scales as the square root of the number of events, so with a factor of $36$ more EMRIs we would measure the slope of the mass function to a precision of $\pm0.05$. This would provide constraints on the low mass end of the MBH mass function that can not be determined in any other way. We note, however, that~\cite{GTV} assumed the dependence of the EMRI rate on MBH mass was known, which is not the case due to the various complications discussed here. LISA will actually determine the product of the MBH mass function with the rate per MBH to the aforementioned precision. Combining EMRI observations with those of MBH mergers could break this degeneracy, but more work is required to properly understand this.

\subsection{Cosmology}
EMRI observations can be used to probe cosmological parameters. A GW observation cannot determine the redshift of an EMRI source and electromagnetic counterparts are unlikely to be observed. However, constraints can also be determined statistically. McLeod and Hogan~\cite{McLeodHogan} showed that by using the electromagnetically determined redshifts of all galaxies within any given LISA EMRI error box, a statistical constraint on the Hubble constant could be determined. A determination of $H_0$ to $1\%$ precision would be possible provided LISA observes $\sim 20$ EMRIs at redshift $z<0.5$. This calculation was done for the classic $5~\mathrm{Gm}$ LISA configuration and assumed that LISA could determine luminosity distance and sky position to precisions $\Delta(\ln D_{L}) =0.07z$ and $\Delta\Omega = 16 z^2$. The new $2.5~\mathrm{Gm}$ baseline will not be able to measure parameters to the same precision. However, these precisions will be reached for some events. In Table~\ref{tab:cosmology} we show how many EMRI events will be observed at redshift $z < 0.5$ in each model and how many will be observed with the assumed parameter-estimation accuracy. We see that most models predict the requisite number of EMRIs, but we would only expect $\sim 5$ events with the assumed precision. We would expect a factor of $2$ worse precision with a factor of $4$ fewer events, and so it is clear that a $2\%$ measurement of $H_0$ will be possible in most cases. The LISA mission is likely to be longer than the two years assumed here, which will increase the number of accurately determined EMRIs. Moreover, all of the EMRIs will contribute something to the $H_0$ measurement, so assuming only the well localised EMRIs are useful is clearly conservative. Thus it is likely that EMRI observations with LISA will provide interesting cosmological constraints, unless the total number of observed EMRIs is toward the low end.

\begin{table}
\begin{center}
\begin{tabular}{c|cc|cc}
&\multicolumn{2}{|c|}{Schwarzschild plunge condition}&\multicolumn{2}{|c}{Kerr plunge condition}\\
Model&$N(z<0.5)$&$N(z<0.5; $ small error$)$&$N(z<0.5)$&$N(z<0.5; $ small error$)$\\\hline
M1&30&5&29&7\\
M2&23&4&22&4\\
M3&62&15&60&16\\
M4&11&4&11&4\\
M5&2&0&3&1\\
M6&35&6&35&8\\
M7&298&48&285&52\\
M8&4&0&4&1\\
M9&25&3&25&5\\
M10&24&0&24&0\\
M11&0&0&0&0\\
M12&354&60&354&74\\
\end{tabular}
\end{center}
\caption{\label{tab:cosmology}Number of EMRIs detected at redshift $z < 0.5$ for each model, computed using each of the two waveform plunge conditions. The first column in each case gives all EMRIs at $z < 0.5$, while the second gives those EMRIs at $z < 0.5$ that also satisfy the error  conditions that were assumed in~\cite{McLeodHogan}, i.e., $\Delta(\ln D_L) < 0.07z$ and $\Delta\Omega < 16 z^2$.}
\end{table}

\subsection{Fundamental physics}
\begin{figure}
\begin{center}
\includegraphics[width=0.5\textwidth]{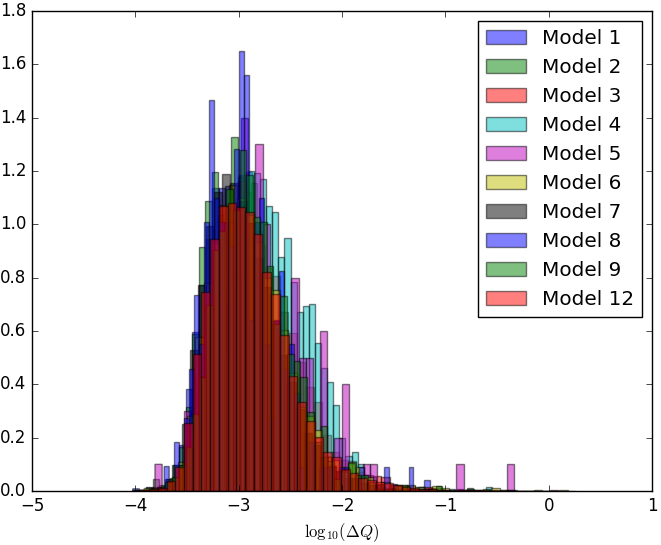}
\end{center}
\caption{\label{fig:SMBHquad}Precision with which LISA could detect a deviation in the quadrupole moment of the MBH spacetime away from the Kerr value.}
\end{figure}

EMRIs are excellent probes of fundamental physics, as they generate many cycles in the strong field regime. In particular, EMRIs can be used to map out the space-time structure outside the central MBH, and identify any deviations from the Kerr metric structure predicted by general relativity. There has been a lot of research in this area and a full discussion of tests of fundamental physics with LISA can be found in~\cite{TestGRLivRev}. An example of the application of EMRIs to fundamental physics is to ask with what precision an EMRI could measure a deviation in the quadrupole moment of a black hole away from the Kerr metric value. Such a study was performed in~\cite{AKBumpy}, using a gravitational waveform model constructed by adding the leading order effect of a quadruple deviation $\Delta Q$ into the AK model~\cite{AK}. We have repeated the same calculation for the current EMRI populations. We assume that EMRIs are occurring in a Kerr MBH spacetime, i.e., the true $\Delta Q = 0$, but compute a Fisher matrix including this quadrupole deviation parameter. The corresponding element of the inverse Fisher Matrix is then a measure of how large a quadrupole deviation could be present in an observed EMRI before we would be able to identify that it was there. These results are shown in Figure~\ref{fig:SMBHquad}, again normalised to a fixed SNR $\rho=20$. We see that every EMRI should provide a constraint on deviations from the Kerr metric at a level of $0.01$--$1\%$, irrespective of our assumptions about the underlying astrophysical model.

\section{Summary}
\label{sec:conc}
We have described a comprehensive study of the prospects for detection of EMRIs with LISA. Our study has attempted to quantify,  for the first time, the uncertainties in EMRI rate predictions arising from astrophysical uncertainties, as well as updating predictions for the new LISA baseline configuration used in~\cite{L3missprop}. We find that LISA should observe several hundred EMRI events over two years, with uncertainties of about one order of magnitude in each direction. These predictions are robust to the distribution of MBH spins and the possible depletion of MBHs at low masses. For all of these events LISA will determine the intrinsic parameters to high precision (sub-percent accuracy), determine sky location to a few square degrees and determine luminosity distance to $O(10\%)$. The parameter-estimation results are largely independent of the astrophysical model assumptions. 

These EMRI detections have tremendous scientific potential. EMRIs will provide new constraints on the mass function and spin distribution of MBHs in the LISA range. EMRIs  will provide constraints on cosmological parameters that are competitive to existing constraints, but with completely independent systematics. Finally, EMRIs will provide strong tests of aspects of the theory of general relativity. We have shown that all of these scientific objectives will be realised irrespective of the true nature of the astrophysical population. The science will be enhanced if the number of observed EMRIs is at the high end of the plausible range, but these goals will still be met even if the rates are nearer to the low end.

\ack AS is supported by the Royal Society. E. Barausse and E. Berti acknowledge support from the H2020-MSCA-RISE-2015 Grant No.\ StronGrHEP-690904. This work has made use of the Horizon Cluster, hosted by the Institut d'Astrophysique de Paris. We thank Stephane Rouberol for running smoothly this cluster for us. E. Berti was supported by NSF Grant No. PHY-1607130 and by FCT contract IF/00797/2014/CP1214/CT0012 under the IF2014 Programme. PAS acknowledges support from the Ram{\'o}n y Cajal Programme of the Ministry of Economy, Industry and Competitiveness of Spain. PAS's work has been partially supported by the CAS President's International Fellowship Initiative.

\end{document}